\documentstyle[epsfig]{mn}

\newcommand\etal{et al.}

\def\wisk#1{\ifmmode{#1}\else{$#1$}\fi}

\def\degree{$^{\circ}$}

\newcommand{\simgt}{\mbox{$\stackrel{>}{_{\sim}}$} }

\newcommand{\kmsmpc}{\,\hbox{km}\,\hbox{s}^{-1}\,\hbox{Mpc}^{-1}}

\newcommand{\be}{\begin{equation}}
\newcommand{\ee}{\end{equation}}

\newcommand{\apj}{{\em Astrophys. J.\ }}

\title[Detection of CMB Structure in a second field with CAT]
{Detection of Cosmic Microwave Background Structure
in a Second Field with the Cosmic Anisotropy Telescope}

\author[J. C. Baker et al.]
{Joanne C. Baker$^{1}$\thanks{New address: Astronomy Department, 
601 Campbell Hall, University of California, Berkeley, CA 94720, USA}, 
Keith Grainge$^{1}$,
M.P. Hobson$^{1}$, 
Michael E. Jones$^{1}$, 
R. Kneissl$^{1}$,
\and
A.N. Lasenby$^{1}$, 
C.M.M. O'Sullivan$^{1,2}$, 
Guy Pooley$^{1}$, 
G. Rocha$^{1,3,4}$, 
Richard Saunders$^{1}$, 
\and 
P.F. Scott$^{1}$, 
E.M. Waldram$^{1}$ \\
    $^{1}$Astrophysics Group,
        Cavendish Laboratory, Madingley Road, Cambridge, CB3 0HE, UK.\\
    $^{2}$Department of Physics, National University of Ireland, 
         Maynooth, Co. Kildare, Ireland. \\
    $^{3}$Department of Physics, Kansas State University, Manhattan, 
          KS~66506, USA. \\
    $^{4}$Centro de Astrofisica da Universidade do Porto, 
          Rua das Estrelas s/n, 4100 Porto, Portugal. }

\date{~}

\pubyear{1999}

\begin{document}
\maketitle


\begin{abstract}

We describe observations at frequencies near 15~GHz of the second 
2\degree $\times 2$\degree\ field imaged with the Cambridge Cosmic 
Anisotropy Telescope (CAT). 
After the removal of discrete radio sources, structure is 
detected in the images on characteristic scales of about half 
a degree, corresponding to spherical harmonic multipoles 
in the range $\ell\approx 330$--680. 
A Bayesian analysis confirms that the signal
arises predominantly from the cosmic microwave background 
(CMB) radiation for multipoles in the lower half of this range; 
the average broad-band power in a bin with centroid $\ell=422$ 
($\theta \approx 51'$)
is estimated to be $\Delta T/T=2.1^{+0.4}_{-0.5} \times 10^{-5}$.   
For multipoles centred on $\ell =615$
($\theta \approx 35'$), we find contamination from Galactic emission
is significant, and constrain the CMB
contribution to the measured power in this bin 
to be $\Delta T/T<2.0 \times 10^{-5}$ ($1\sigma$ upper limit).  
These new results are consistent with the first detection made by
CAT in a completely different area of sky.
Together with data from other experiments, this new CAT detection 
adds weight to earlier evidence from CAT for a downturn in the 
CMB power spectrum on scales smaller than 1\degree.  Improved 
limits on the values of $H_0$ and $\Omega$ are determined using the new
CAT data. 

\end{abstract}



\newpage
\section{Introduction}

Observations of spatial fluctuations in the cosmic microwave background (CMB)
radiation are fundamental to our understanding of structure formation 
in the universe as they mark the earliest observable imprints 
of massive gravitational structures (see e.g. review by White, Scott \&
Silk 1994). The distribution and amplitude
of anisotropies in the CMB sky over scales from degrees to arcminutes can 
be used to discriminate between competing cosmological theories. 
On scales of 0.2\degree - 2\degree, inflationary 
models predict that increased power should be seen in the CMB sky due to 
scattering of photons during acoustic oscillations of the photon-baryon 
fluid at recombination. Detection and study of these acoustic or 
`Doppler' peaks 
in the power spectrum is one of the primary goals of CMB astronomy. 
Furthermore, the amplitudes and angular scales of the acoustic peaks 
provide powerful constraints on basic cosmological parameters
including $H_0$ and $\Omega$. 

The first clear indication of a downturn in the power spectrum on
sub-degree scales was provided by the detection of CMB power
by the Cambridge Cosmic Anisotropy Telescope (CAT)
on scales of about half a degree (Scott et al. 1996; Paper I). 
The CAT is a three-element interferometer operating at frequencies
between 13 and 17~GHz (Robson et al. 1993). It is sensitive to structure
on angular scales of about $10'$ to $30'$ over a field of view
covering 2\degree$\times 2$\degree\ (primary beam FWHM). This 
paper describes observations of a second field observed with CAT and 
the detection of CMB anisotropies within it, at levels consistent
with measurements in the first field.

\section{Observations and Data Reduction}

Observations have been made with the CAT of a blank field centred at the 
position 17~00~00 $+64$~30~00 (B1950), which we
call `CAT2'. The field was chosen to be relatively free from strong
radio sources at frequencies up to 5~GHz (Condon, Broderick \& Seielstad 
1989), lie at high Galactic latitude ($b>30$\degree) and away from 
known Galactic features (e.g. the North Polar spur). 

In periods during the interval 1995 March -- 1997 June, CAT observed
the CAT2 field at three frequencies, 13.5, 15.5 and 16.5 GHz.  The 
three baselines of the array were scaled with frequency to achieve the
same resolution at each frequency. The resulting synthesised beams 
measured $20'$ FWHM in right ascension and $24'$ in declination.
The primary beam of the telescope, due to the symmetric 
nearly-Gaussian envelope
beams of the three horn-reflector antennas, has a FWHM of 1.96\degree\ at
15.5~GHz, scaling inversely with frequency. The telescope observes in 
two orthogonal linear polarisations (which rotate on the sky as it 
tracks), and 
has a system noise temperature of 50~K. An observing bandwidth of 500~MHz
was used. Amplitude and phase calibrations were carried out daily with 
observations of Cas~A (using the flux scale of Baars et al. 1977), 
and cross checked periodically by observations of other 15-GHz 
calibrators from the VLA list (Perley 1982). Typical uncertainties 
in flux scaling are less than 10\%. Observations
were generally carried out at night (about 80\% of the data)
or pointing $>90^{\circ}$ away from the sun to avoid possible solar 
interference; no extra emission from the moon was detected at 
this declination.

The CAT2 data were reduced using the same method as for the first field, CAT1,
as described in detail by O'Sullivan et al. (1995). Phase rotation and
flux calibration were applied first, and the 
data were then edited and analysed using standard tasks in AIPS. 
Excessively noisy data (reflecting periods of poor
weather) were excised --- the visibility amplitudes
were all checked by eye for periods where they 
regularly exceeded the mean value by more than 3$\sigma$ and then these 
data ($\pm 1$ hour) were removed from all baselines. Across 
the remaining dataset, individual visibilities with amplitudes exceeding 
the 3$\sigma$ threshold were also excluded. In total, about 40\% of 
the data were excluded by this process
leaving 370, 310 and 1340 hours of good data at 13.5, 15.5 and 16.5~GHz
respectively. Since the atmospheric coherence time is very short (10s)
compared with the total integration time, any remaining
atmospheric signals will be distributed uniformly across the synthesised
map as noise, unlike true sky signals which will be modulated by the envelope
beam pattern. The efficacy of atmospheric filtering for the CAT interferometer 
has already been demonstrated (Robson et al. 1994). CAT's sidelobe response
and lack of crosstalk and correlator offsets are discussed in
O'Sullivan et al. (1995). No radio interference was seen.

\section{Source Subtraction}

Radio sources contributing significantly to the CAT2 image were identified
and monitored at 15~GHz using the Ryle Telescope (RT). 
Five RT antennas were used in a configuration giving a synthesised beam 
of $30''$ FWHM. The RT has an instantaneous field of view of $6'$ FWHM,
but for these observations is used in a rastering mode which covers $30'
\times 30'$ in 12 hours to a typical flux sensitivity of
1.5~mJy/beam. To detect sources within the the CAT2 field, the central
2\degree$\times 2$\degree\ area was scanned with the RT in raster mode
in sixteen days.  A source list was then compiled, including sources
listed in the Green Bank 4.85-GHz survey (Condon et al. 1989). Pointed
observations with the RT were then made, and repeated regularly to
check for variability over the whole period of the CAT
observations. In all, twenty-nine sources were detected, the faintest
having a flux density of 4.5~mJy at 15~GHz.  Flux densities at 13.5
and 16.5~GHz were extrapolated using spectral information obtained
from lower frequency surveys (i.e. 4.85~GHz, Condon et al. 1989, and
1.4~GHz, Condon et al. 1998) where available.  A flat spectrum between
13.5 and 16.5~GHz was assumed for three sources without other
data. After correcting for CAT primary beam attenuation, the
corresponding flux densities were subtracted as point sources from the
visibility data.  Totals of 33\,mJy at 13.5, 27\,mJy at 15.5 and
20\,mJy at 16.5 GHz were subtracted. The robustness of the
source-subtraction procedure is illustrated in O'Sullivan et al. The
strongest source in the field, at position 16~45~32~+63~35~29 (B1950),
was highly variable (by factors of up to two over periods of a few
days) and so was monitored regularly with the RT; no residuals at the
source position remain after subtraction of the variable source. The
successful subtraction of this source, 1\degree away from the pointing
centre and clearly visible at the same position in all three maps,
shows that CAT phase calibration and pointing are accurate.

\section{The Images}

The resulting source-subtracted images each show excess signal in the 
central 2\degree$\times 2$\degree\, falling away at larger radii as
expected from  the antenna envelope beam. These are displayed in Figure
\ref{cat2ss}.  The instrumental noise levels 
(measured directly from the visibilities) for each source-subtracted
image are 6.1, 6.5 and 3.5 mJy/beam rms for 13.5, 15.5 and 16.5 GHz
respectively.  At the three frequencies, excess powers above the
intrinsic noise level in the central 2\degree$\times 2$\degree\ area
of $7.8 \pm 1.0$, $8.2 \pm 1.0$ and $8.3 \pm 0.5$ mJy/beam rms were
found for the source-subtracted 13.5, 15.5 and 16.5~GHz images
respectively. Checks were made by splitting the data in time
(consistent excesses were seen), polarisation (no excess power was
visible on polarisation difference maps) and cross-correlating the
source-subtracted image with a reconstructed map of the radio sources
as in O'Sullivan et al. (no correlation was found); most importantly,
maps were made by correlating orthogonal polarisations and these gave
the {\em same} noise levels as those above.

To attempt to remove the instrumental response and thereby illustrate 
the distribution of features on the sky, we have co-added the data 
from the three frequencies weighted as $\nu^{2}$ and CLEANed the final 
image. This is shown in Figure \ref{cat2cln} and shows 
the presence of significant features in  the central region 
as well as the diminution of the telescope 
sensitivity in accordance with the envelope beam.

The strongest feature in the CAT2 images is a negative one
centred at position 17~05~29~+64~47~37 (B1950) reaching $-39$ mJy 
at 16.5 GHz (i.e. 5$\sigma$, relative to the rms excess power
in the sky).  For comparison, the strongest positive feature in the
16.5-GHz dirty map has a peak flux density of 26 mJy at position 
16~55~55 +63~49~47 (B1950). 
The negative feature can be seen at all three frequencies 
(most clearly in the 13.5 and 16.5-GHz images in Figure \ref{cat2ss}), 
both before and after source subtraction and in 
different time cuts, and its spectrum is consistent with that of 
CMB radiation.  For example, even before source subtraction
the hole reaches $-30$ mJy in the 16.5-GHz image, and only three
weak sources ($S_{16.5} < 10$ mJy) lie within $40'$ of it. No obvious 
Galactic structures at the position of the negative feature were 
visible in IRAS 100\,$\mu$m (Wheelock et al. 1994)
or H\,{\sc i\/} 21-cm sky survey images (Lockman \& Dickey 1995) or
the $100\mu$m map of Schlegel, Finkbeiner \&
Davis 1998. Indeed, there is no resemblance at all between any of
these images of the CAT2 region and the CAT2 image itself.

On the scales sampled by CAT, the negative feature is 
unlikely to have been caused by a Sunyaev-Zel'dovich 
(S--Z) effect (Sunyaev \& Zel'dovich 1972; Rephaeli 1995)
towards a single massive cluster. Any cluster which might produce
such an effect would have to be nearby (subtending a large
angle on the sky, filling the CAT beam) and/or very massive
to produce such a strong signature. The strongest 
S--Z decrements measured towards nearby clusters 
at 15~GHz with the RT (Grainge \etal\ 1996; Jones \etal\ 1993)
are about $-0.5$\,mJy on arcminute scales. Observing a similar cluster
at $z \sim 0.2$ (e.g. gas mass $10^{14} M_{\odot}$, with a King profile 
of core radius $\sim 300$ kpc, truncated at about ten core radii) with the 
larger CAT beam of $30'$ 
would produce a similar decrement amplitude, due to the balance 
between the effects of beam dilution and sampling cluster gas out to 
a larger radius. In order to maximise the S--Z signal in a $30'$ beam 
with the minimum gas mass, a nearby cluster ($z \approx 0.05$) with
$10^{15} M_{\odot}$ of gas would be required, which is patently not 
observed. A system at any higher redshift would require an even 
higher mass because of beam dilution.

A central portion of the CAT2 field has been observed serendipitously 
with  ROSAT. The direction of the negative CMB feature lies close to 
the edge of the PSPC field ($45'$ off axis) and is partially shadowed 
by the support structure of the PSPC detector. An X-ray source 
(heavily distorted by the PSPC point spread function) is found 
nearby (at the position 17~05~49 +64~45~50, B1950) but its X-ray spectrum 
does not fit a thermal model for any reasonable temperature, and 
is fit much better by a power law, implying it is not a cluster (but
perhaps an AGN or Galactic source).  Four clusters (Abell 2246 and 
two others at $z = 0.25$ plus one at $z = 0.44$) --- and an optically 
luminous QSO (HS~1700+6416) ---  have been detected by ROSAT and lie
within the central $10'$ of the CAT2 field (Reimers et al. 1997; 
Vikhlinin et al. 1998). 
All four clusters are fainter than $10^{44}$ erg\,s$^{-1}$ at X-ray
energies 0.4--2.0 keV and none is apparent in the CAT images.

Finally, we emphasise that the presence of a 5$\sigma$ feature in a 
single CAT field should not be construed as evidence for non-Gaussianity 
of the CMB fluctuations: the sidelobe structure of the synthesised beam
of the three-element CAT is significant, and analysis should be carried
out  in the aperture plane --- see Section \ref{analysis}. As a check, we have
simulated CAT images given standard CDM-based realisations of CMB
structure (using the actual CMB and Galactic mean power measured by 
CAT), and find that features which appear as strong as $5\sigma$ occur 
in about 10\% of cases. We return to this point in Section \ref{analysis}.

\section{Determination of the CMB Component}
\label{analysis}

Due to the limited range of baselines and resulting 
sparse sampling of the $u-v$ plane by CAT, a statistical likelihood 
analysis was employed to estimate the relative contributions of CMB 
and Galactic components given the three-frequency CAT data. 
A Bayesian likelihood method was used, as in Paper I, as described by Hobson, 
Lasenby \& Jones (1995). This method uses the complex visibility data 
directly in the calculation of the likelihood function to avoid 
the problem of the long-range correlations present between different
resolution elements in the image plane.

As described in Paper I, power was estimated in two 
independent annular bins centred on spherical harmonic multipole 
values of $\ell =410$ and $\ell =590$ and with widths equivalent to 
the diameter of the antenna function, thus together spanning the range 
$\ell\approx 330-680$. The noise-weighted centroid positions for 
data in each bin are $\ell =422$ and $\ell =615$. CMB and Galactic signals 
were modelled as independent Gaussian distributions with a power-law 
spectrum ($S \propto \nu^{\alpha}$), the CMB with a fixed spectral index 
of $+2$ in flux density and the Galactic spectral index variable 
between 0 and $-1$ (as expected for Galactic free-free and synchrotron 
emission).

After  marginalising over the Galactic parameters, this analysis 
confirmed that the bulk of the power in the 16.5~GHz map (Fig. \ref{cat2ss})
arises from CMB fluctuations. 
The CMB component was clearly distinguished 
from possible Galactic contamination in the
$\ell =422$ bin; $\Delta T/T = 2.1^{+0.4}_{-0.5} \times 10^{-5}$ was 
estimated for the CMB signal, compared with only 
$\Delta T/T = 0.8^{+0.5}_{-0.8} \times 10^{-5}$ for any Galactic
component. The uncertainties quoted are $1\sigma$ values. 
The CMB--Galaxy separation was less certain in the $\ell =615$ bin ---
a Galactic contribution of $\Delta T/T = 1.0^{+0.7}_{-1.0}
\times 10^{-5}$ was given by the likelihood analysis, which is
equivalent to an upper limit for the marginalised CMB power of 
$\Delta T/T < 2.0 \times 10^{-5}$ ($1\sigma$).  
These values are plotted in Figure \ref{deltat}. 
The average values of $\Delta T/T$ in the CAT2 field agree 
with the CAT1 result (Paper I) within $1\sigma$; for comparison the
measured CMB powers in the CAT1 field were  
$\Delta T/T =1.9^{+0.5}_{-0.5} \times 10^{-5}$  ($\ell = 420$) and 
$\Delta T/T =1.8^{+0.7}_{-0.5} \times 10^{-5}$  ($\ell = 590$).

We have also investigated how much of the CMB power is not associated
with the strong negative feature discussed in Section 4. We removed
the dip as a point source from the visibilities and repeated the
above analysis. Significant power remained, at around half of the
level with the negative feature included.

\section{Estimation of Cosmological Parameters using new CAT results}

The CAT2 points, taken in conjunction with the results from the Saskatoon
experiment \cite{sask}, provide further evidence for a downturn in
the CMB power spectrum for $\ell \simgt 300$. To assess the
implications for cosmological parameters using current CMB data
sets, we have extended the analysis presented previously by
Hancock \etal\ (1998). This analysis used a statistically
independent subset of the current data and carried out $\chi^2$
fitting for a range of cosmological models and parameters. The
extensions carried out for the present work were (a) the inclusion
of the CAT2 point at $\ell = 422$; (b) the inclusion of new points
from experiments Python (Python III, Platt \etal\ 1997), 
MSAM (the 2nd and 3rd flights, Cheng \etal\ 1996, 1997, Ratra
\etal\ 1997), ARGO (Aries+Taurus region, Masi \etal\ 1996),
FIRS \cite{firs} and BAM \cite{bam}, and with the latest calibration
correction to the Saskatoon data (i.e. increased by 5\%, Leitch, 
private communication); (c) consideration of a wider class of 
cosmological models, all treated
using exact power spectra rather than the generic forms assumed
in Hancock \etal\ (see also Rocha 1997; Rocha \etal\ in preparation).
The formalism and approach are otherwise the same as in
Hancock \etal\ (1998), to which the reader is referred.
The cosmological models considered were:
\begin{enumerate}
\item Flat models with $\Lambda=0$, and with a range of spectral tilts 
(i.e. $n \neq 1$);
\item Flat models with $\Lambda \neq 0$, and a range of spectral
tilts;
\item Open models with $\Lambda=0$ and open-bubble inflation spectrum 
\cite{sugi,ratra2,kami}.
\end{enumerate}
In cases (i) and (ii), nucleosynthesis constraints, with 
$0.009 \leq \Omega_{b} h^{2} \leq 0.02$
\cite{copi} were assumed.  Theoretical power spectra from Seljak \& 
Zaldarriaga (1996) were used in cases (i) and (ii), and kindly provided by
N. Sugiyama for case (iii). 
The parameters fitted for (unless fixed) were the
Hubble constant $h$ (in units of $100 \kmsmpc$ ), the spectral
tilt $n$, the cosmological constant $\Omega_{\Lambda}$ and the matter 
density $\Omega_m$ in units of the critical
density. The flat models are defined by 
$\Omega_m + \Omega_{\Lambda} = 1$. 
Note the ranges of $h$ and $\Omega_{\Lambda}$ considered were $0.3 - 0.8$
and $0.3 - 0.7$ respectively.
 
The results are displayed in Table~\ref{tabum}. In each case the best fit
values of the parameters are shown, together with marginalised
error ranges. These marginalised errors are $\pm 1\sigma$
confidence limits formed by integrating over all the other 
parameters with a uniform prior.

\begin{table}
\begin{center}
\begin{tabular}{llccc}
\multicolumn{1}{l}{Model}&\multicolumn{1}{l}{}&\multicolumn{1}{l}{Best Fit}&\multicolumn{2}{l}{Marginalised ranges}\\ 
		&          &value	& ($1\sigma$) & ($2\sigma$)	     \\
Flat models     &$h$ & 0.3 	        &0.3 -- 0.55    &0.3 -- 0.8 \\
		&$n$ & 0.88		&0.8 -- 1.15    &0.65 -- 1.3 \\ 
		&&			&		& \\		
Lambda models   &$\Omega_{\Lambda}$ &0.3&0.4 -- 0.7     &0.3 -- 0.7 \\
		&$h$ &0.35      	&0.3 -- 0.55    &0.3 -- 0.8  \\ 
		&$n$ &0.9		&0.8 -- 1.1     &0.7 -- 1.2   \\
		&&		         &		& \\	
Open models     &$\Omega_m$ &0.7	 &0.5 -- 1.0    &0.3 -- 1.0 \\
		&$h$ &0.6                  &0.3 -- 0.7  &0.3 -- 0.8  \\ 
		&&			 &		& \\	  	
\end{tabular}
\caption{Results of the fitting analysis. Non-stated parameters were
marginalised over to estimate the $1\sigma$ and $2\sigma$ ranges. 
 }
\label{tabum}
\end{center}
\end{table}

In common with other recent work on fitting to current CMB data
(e.g. Lineweaver \& Barbosa 1998) a tendency to low $H_0$ values 
(except in the case of open
models) is found, although it is clear from the marginalised
ranges that the statistical significance of this is not yet high.
For comparison with the results of Webster \etal\ (1998), who worked
jointly with recent CMB and IRAS large-scale structure results, we
note that a flat $\Lambda$ model with normalization fixed to the
COBE results and $n=1$ yielded a best fit of
$\Omega_{\Lambda}=0.7$ and $h=0.6$.  In conjunction with the CAT
points, new results forthcoming from the Python, Viper and 
QMAP\footnote{Results from QMAP published after this 
paper was submitted show a rise in the power spectrum from 
$\ell\sim 40$ to $\ell \sim 200$ consistent with
the Saskatoon results (de Oliviera Costa et al. 1998).}
experiments may soon be very significant in ruling out open
models and in further delimiting the Doppler peak, sharpening up
these parameter estimates. 
Recent OVRO results at $\ell \sim 590$ (Leitch, private
communication) agree well with the values found by CAT. The joint 
CAT and OVRO results will clearly be very significant in 
constraining the latest cosmic string power spectrum predictions, 
which now include a cosmological constant (Battye \etal\ 1998). 
These predictions succeed in recovering a significant `Doppler peak' 
in the power spectrum, but have peak power for $\ell$ in the range 500 to
600, at variance with the trend of the current experimental results.

\section{Conclusions}

\begin{itemize}

\item[(1)]
We have imaged a 2\degree $\times 2$\degree\ patch of sky 
at 13-17 GHz with the CAT; this is the second field observed by CAT.

\item[(2)] 
Significant CMB anisotropy is detected in the CAT2 field with an average
power of $\Delta T/T = 2.1^{+0.4}_{-0.5} \times 10^{-5}$ for 
the $\ell =422$ bin, and with an upper limit of
$\Delta T/T < 2.0 \times 10^{-5}$ for $\ell =615$ 
(due to Galactic contamination).

\item[(3)] This new result is consistent with the 
first detection made by CAT in a different area of sky. 

\item[(4)] Together with other CMB data over a range of angular scales, the
inclusion of the new CAT2 detection restricts the likely values of 
cosmological parameters.

\end{itemize}

\section*{Acknowledgments} 
Staff at the Cavendish Laboratory and MRAO, Lords Bridge are 
thanked for their ongoing support in the running of CAT. We thank an
anonymous referee for comments.
We are pleased to acknowledge major PPARC support of CAT.
G. Rocha wishes to acknowledge a NSF grant EPS-9550487 with
matching support from the State of Kansas and from a K$\ast$STAR 
First award. 
We also thank N. Sugiyama, and U. Seljak and M. Zaldarriaga
for access to their theoretical power spectra. 



\begin{figure}
\caption[fig]{ Images of source-subtracted CAT2 data 
at a) 13.5, b) 15.5 and c) 16.5 GHz. RA and Dec (B1950) 
and grey scale flux ranges (mJy/beam) are indicated. 
Instrumental noise levels are 6.1, 6.5 and 
3.5 mJy/beam rms for a), b), c) respectively, reflecting
the total integration times. 
}
\label{cat2ss}
\end{figure}

\begin{figure}
\caption[fig]{ CLEANed image of co-added CAT2 data (three frequencies
weighted as $\nu^{2}$.)
  }
\label{cat2cln}
\end{figure}

\begin{figure}
\caption[fig]{Plot of average marginalised power ($\ell^{2} C_{\ell} / 2 \pi$
against $\ell$) in 
CMB fluctuations (solid lines) and Galactic emission (dotted lines)
in the CAT2 field as estimated by the Bayesian method 
for the two ($\ell =422$ and $\ell =615$) bins. 
Error bars ($1\sigma$) are shown. The prediction of a standard CDM model
($\Omega_{m} = 1$, $\Omega_{\Lambda} = 0$, $\Omega_{b} = 0.05$; 
$h=0.5$,$n=1$, no tensors) is shown (solid curve) for comparison. 
  }
\label{deltat}
\end{figure}

\end{document}